\documentclass[13pt,a4paper,oneside]{article}
\usepackage[utf8]{inputenc}
\usepackage{relsize}
\usepackage[T1]{fontenc}         
\usepackage{authblk}
\usepackage{amsmath}             
\usepackage{amssymb}             
\usepackage{amsfonts}            
\usepackage{latexsym}            
\usepackage{amsthm}              
\usepackage{braket}
\usepackage{graphicx}
\usepackage{float}              
\usepackage{tabularx}
\usepackage{cancel}
\usepackage{aligned-overset}

\usepackage{hyperref}

\usepackage{algorithm}
\usepackage{algpseudocode}

\usepackage{tikz}
\usetikzlibrary{spy}
\usepackage{xcolor}
\usepackage{listofitems}

\usepackage{caption}
\usepackage{subcaption}

\algnewcommand{\Initialize}[1]{%
  \State \textbf{Initialize:}
  \Statex \hspace*{\algorithmicindent}\parbox[t]{.8\linewidth}{\raggedright #1}
}

\newcommand{\R}{\mathbb{R}}

\newcommand{\oper}[1]{\operatorname{\mathcal{#1}}}
\newcommand{\RecOp}{\oper{R}}
\newcommand{\FwdOp}{\oper{A}}
\newcommand{\NNOp}{\Lambda}
\newcommand{\NNparam}{\theta}
\newcommand{\signal}{f}
\newcommand{\data}{g}
\newcommand{\RecSpace}{X}
\newcommand{\DataSpace}{Y}

\DeclareMathOperator*{\argmin}{arg\,min}

\theoremstyle{plain}

\newtheorem{proposition}{Proposition}[section]

\theoremstyle{definition}

\theoremstyle{remark}

\numberwithin{equation}{section}

\title{Enabling self-supervised learned primal dual with Noise2Inverse}
\author[1]{Antti Sällinen}
\author[3]{Siiri Rautio}
\author[1]{Santeri Kaupinmäki}
\author[1,2]{Andreas Hauptmann}
\affil[1]{Research Unit of Mathematical Sciences, University of Oulu, Finland}
\affil[2]{Department of Computer Science, University College London, United Kingdom}
\affil[3]{Department of Mathematics and Information Science, Josai University, Japan}

\begin{document}
\maketitle

\begin{abstract}
X-ray computed tomography reconstruction is an ill-posed inverse problem, particularly in low-dose and sparse-angle settings where measurements are noisy and incomplete. While learned reconstruction methods such as the Learned Primal-Dual algorithm achieve strong performance, they typically rely on supervised training with access to ground-truth data, which is often unavailable in practice.

In this work, we propose a self-supervised reconstruction method by extending the Noise2Inverse framework to the Learned Primal-Dual algorithm. The resulting approach, called Noise2Inverse Learned Primal-Dual (N2I-LPD), enables training of a learned iterative reconstruction operator without ground-truth images by exploiting the statistical independence of noise in distinct measurements with respect to angular rotation of the CT-scan. 

We compare the proposed method with classical reconstruction methods, as well as neural network–based approaches such as a U-Net trained within the same N2I framework. The results demonstrate that N2I-LPD achieves improved reconstruction quality, highlighting the potential of combining learned reconstruction operators with self-supervised training strategies for practical CT imaging scenarios where ground-truth data is unavailable.
\end{abstract}

\section{Introduction}
X-ray computed tomography (CT) is a widely used imaging modality in medical and industrial applications due to its ability to provide high-resolution cross-sectional reconstructions of objects. Mathematically, CT reconstruction can be formulated as an inverse problem where the goal is to recover an unknown object from its line integral measurements \cite{kak1988principles,natterer1986mathematics}. The inverse problem is ill-posed, particularly when the number of projections is limited, and becomes increasingly challenging in low-dose settings due to high noise levels \cite{ chen2017low,fessler2000statistical}.

In scenarios where projection data is densely sampled over a full angular range and acquired with sufficiently high radiation dose, analytical reconstruction methods such as filtered backprojection (FBP) \cite{natterer1986mathematics} produce reconstructions of sufficient quality. However, if these conditions are not met, the reconstruction task becomes significantly more challenging. In low-dose settings, the measurement is corrupted by increased noise, which is directly propagated to the reconstruction by FBP, resulting in degraded image quality. In sparse-angle settings, where only a limited number of projections are available, the data becomes incomplete and the inverse problem becomes more ill-posed. As a result, reconstructions suffer from artifacts.

To address these challenges, reconstruction methods based on variational regularization and iterative optimization have been widely studied. In classical approaches, the reconstruction is formulated as a regularized optimization problem, for instance, using Total Variation (TV) \cite{rudin1992nonlinear}, and solved with iterative methods such as the Primal-Dual Hybrid Gradient (PDHG) algorithm \cite{chambolle2011first,sidky2012convex}. In the past decade, machine learning approaches have emerged as powerful tools for CT reconstruction. These data-driven methods leverage neural networks to improve reconstruction quality and are now considered state-of-the-art in many settings \cite{arridge2019solving,wang2018image,wang2020deep}.

One of the simplest ways to incorporate neural networks into CT reconstructions is through post-processing \cite{arridge2019solving,wang2020deep}. In this approach, some knowledge-driven reconstruction is first applied to the measured data to obtain an initial reconstruction, which is then processed by a neural network. The network aims to remove noise and artifacts from the initial reconstruction. These methods are relatively easy to train and implement, but they typically require large amounts of training data. Convolutional Neural Networks (CNNs), particularly U-Net–based architectures, have been shown to perform well in this setting \cite{jin2017deep,kang2017deep}.

An alternative approach is to learn the reconstruction operator as an unrolled iterative scheme \cite{adler2017solving,adler2018learned}. In these methods, neural networks are embedded into iterative reconstruction schemes, allowing the entire reconstruction process to be learned from data. A prominent example is the Learned Primal-Dual (LPD) algorithm \cite{adler2018learned}, which unrolls a primal-dual optimization method into a trainable architecture.

Most data-driven reconstruction methods are trained in a supervised setting, where paired data consisting of inputs (measurements or preliminary reconstructions) and corresponding ground-truth images are required. In CT imaging, this would require both a low-dose and a high-dose, densely sampled scan, where the latter serves as data for a reference reconstruction. However, in clinical practice, acquiring such data is infeasible, as it would expose patients to unnecessary levels of ionizing radiation by multiple and high-dose scans. This lack of ground-truth data constitutes a major limitation for supervised learning approaches in medical imaging.

This limitation of supervised learning approaches motivates the use of self-super\-vised learning methods, which enable training without ground-truth images by exploiting the statistical structure of the measurement noise. In recent years, several such methods have been proposed for imaging applications including Noise2Noise \cite{lehtinen2018noise2noise}, Noise2Self \cite{batson2019noise2self}, Noise2Void \cite{krull2019noise2void}, and specifically for CT imaging as Noise2Inverse \cite{hendriksen2020noise2inverse}. Even more recent approaches exist, including Equivariance2Inverse \cite{schut2025equivariance2inverse} and Noisier2Inverse \cite{gruber2025noisier2inverse}. These approaches typically rely on the assumption that noise is mean-zero and statistically independent across measurements.

In this work, we extend the self-supervised Noise2Inverse (N2I) framework, which utilizes the FBP, to the Learned Primal-Dual reconstruction method. This results in a self-supervised unrolled learned reconstruction operator that does not require ground-truth images during training. We evaluate the proposed method on simulated data as well as experimental data to test generalizability. For comparison, we train U-Net models both in a supervised manner and in a self-supervised manner using the Noise2Inverse framework. Additionally, we investigate the effect of the rectified linear unit (ReLU) activation function on training and reconstruction quality.

The paper is organized as follows. In section \ref{sec:tomography}, we introduce the inverse problem of X-ray tomography and explain how CT scanning is modeled mathematically. Section \ref{sec:methods} recalls the Noise2Inverse framework, the Learned primal-dual method, and introduces how to combine them to train the learned reconstruction operator. In section \ref{sec:experiments}, the experimental setup is discussed, including introducing the data sets and explaining the practical neural network implementation details. The results are presented in section \ref{sec:Results}. The section \ref{sec:discussion} is for discussion, where we lay out the implications of our findings and potential future research. Finally we have the section \ref{sec:conclusion} for conclusions, where we shortly state the final remarks of this work.

\section{X-ray tomography} \label{sec:tomography}
The task of CT reconstruction can be framed as an inverse problem \cite{natterer1986mathematics,kak1988principles}:
\begin{equation}\label{eq:inverse_problem}
 \data = \FwdOp\signal + \varepsilon,
\end{equation}
where $\data$ denotes the measured X-ray data (sinogram), $\FwdOp\colon\RecSpace\to\DataSpace$ is the X-ray transform describing the measurement process and geometry, and $\varepsilon$ represents measurement noise which is assumed to be zero-mean Gaussian. The inverse problem \eqref{eq:inverse_problem} is ill-posed and due to real-world complications such as incomplete sampling and measurement noise, regularization is required to obtain stable reconstructions \cite{engl1996regularization}.

Here, we will use the X-ray transform to model a fan-beam geometry. It is an integral transform closely related to the Radon transform and integrates the attenuation coefficient along lines through the object \cite{natterer1986mathematics}. The measurement data in CT are obtained by collecting line integrals of the object from multiple projection angles.
The X-ray transform is defined as follows. Let $\signal \in \RecSpace = L^2(\R^n)$, we define lines $L\subset\R^n$ with directions defined by angles $\phi \in S^{n-1}$, incident at detector locations $s\in \R^n$, such that the line coordinates $x \in L(\phi, s)$ are parametrized by $t \in \R$ as $x(t) = s - t\phi$. We then have that the X-ray transform $\FwdOp: X \to Y$ is defined by the line integral
\begin{equation}\label{eq:x_ray_trafo}
 \FwdOp\signal(\phi, s) =  \int_{x\in L(\phi,s)}f(x)\,\mathrm{d}x = \int_{-\infty}^\infty f(s-t\phi)\,\mathrm{d}t.
\end{equation}

In a CT scanner, X-ray beams are emitted with initial intensity $I_0$, they pass through the object, and are detected with reduced intensity $I_1$. According to the Beer--Lambert law \cite{natterer1986mathematics}, the logarithmic attenuation measured by the detector corresponds to the line integral of the attenuation coefficient:
\begin{equation}\label{eq:BLint}
    \ln \frac{I_0}{I_1} = \int_{-\infty}^\infty f(s-t\phi)\,\mathrm{d}t.
\end{equation}

This relationship forms the basis for modeling the measurement process and analyzing measurement noise. In practice, CT scanners acquire only a finite number of such measurements from different projection angles. Consequently, in practice the forward operator used in reconstruction algorithms corresponds to a discretized version of the X-ray transform. 

In this work, we consider the fan-beam geometry, which is a two-dimensional scanning setup where a point source emits X-rays that propagate through the object and are detected by a detector array. During acquisition, the source-detector system rotates around the object, collecting projections at multiple angular positions and producing the sinogram $y$. 

More precisely, this setting is defined with two variables: angle increments and the amount of X-rays. The angle increments give one dimension of the sinogram. Those increments tell us how many angular samples the sinogram has. The second dimension is given by the amount of X-rays, and it informs the density of the detected information. To add, angle increments act as an independent variables, which is necessary to know for Noise2Inverse later in this paper.

The goal of the inverse problem is to recover $x$ from the measurements in \eqref{eq:inverse_problem}. There are many approaches to solving this problem, ranging from analytical reconstruction methods to variational regularization and modern machine learning techniques.

One classical approach is the Filtered Back Projection (FBP) algorithm, which approximates a filtered inverse of the X-ray transform in \eqref{eq:x_ray_trafo} \cite{natterer1986mathematics}. The filtering operation is defined as
\begin{equation*}
    \mathcal{G} g(\phi, s) = \frac{1}{2\pi} \int_{-\infty}^\infty (\mathcal{F}g)(x) |r| e^{i r t} \ \mathrm{d}r.
\end{equation*}
The FBP algorithm reconstructs the object by first filtering the projection data in the frequency domain as above and then backprojecting the filtered projections over the image domain. The reconstruction formula  is then given by 
\begin{equation}\label{eq:FBP_algo}
    f \approx  \FwdOp^\dagger g := \FwdOp^*\mathcal{G} g = \FwdOp^*\mathcal{G} \FwdOp f ,
\end{equation}
where $\FwdOp^*$ is the adjoint of the forward operator $\FwdOp$ in certain geometry.

The FBP method provides very fast reconstructions, but assumes densely sampled projection data covering the full angular range. When these assumptions are violated, for example in sparse-angle settings, the resulting reconstructions contain artifacts. Furthermore, in the presence of noisy measurements, such as in low-dose CT imaging, the reconstructed images are typically noisy since the FBP algorithm does not incorporate explicit denoising or regularization.

An alternative classical approach is variational regularization, which incorporates prior information to stabilize the reconstruction. This is particularly useful in ill-posed situations, such as when the data are noisy or incomplete. The inverse problem \eqref{eq:inverse_problem} can be formulated as the minimization problem
\begin{equation}\label{ref:variational}
  \argmin_f \mathcal{E}_\alpha(f) = \argmin_f ||\FwdOp f - \data||_2^2 + \alpha || |\nabla f |||_1,
\end{equation}
where $||\cdot||_2^2$ is the data discrepancy term and $||\cdot||_1$ is the regularization functional with regularization parameter $\alpha$ controlling the trade-off between data fidelity and regularization.

In this work, we consider Total-Variation (TV) regularization \cite{rudin1992nonlinear}, as written above.

Many modern reconstruction algorithms can be interpreted as iterative schemes for solving variational problems of the general form of \eqref{ref:variational}. In recent years, learning-based approaches have been proposed that replace parts of these iterative algorithms with trainable neural network components. This leads to learned reconstruction schemes that combine the structure of classical optimization methods with the expressive power of deep learning. Such learned reconstruction operators can be designed so that they are able to solve the inverse problem \ref{eq:inverse_problem}.

Let us consider a family of parameterized mappings
\begin{equation*}
    \mathcal{R}_\NNparam: Y \to X,
\end{equation*}
called learned reconstruction operators \cite{hauptmann2025learnediterativenetworksoperator}, where $\NNparam \in \Theta$ is a specific parameter configuration obtained from the parameter space $\Theta$ via a training procedure. The goal is to provide suitable training data, both in quality and amount, in order to train the operator $\mathcal{R}_\NNparam$, solving the inverse problem as
\begin{equation*}
    f = \mathcal{R}_\NNparam(g).
\end{equation*}

Post-processing methods, which pair FBP reconstruction with a neural network, first apply an analytical reconstruction operator $\mathcal{A}^\dagger$ to obtain an initial reconstruction, which is then processed by a neural network $\Lambda_\theta$. The network takes the initial reconstruction as input and outputs a denoised or artifact-reduced image. This kind of learned reconstruction operator can be expressed as $\mathcal{R}_\NNparam = \NNOp_\NNparam \circ \FwdOp^\dagger$. Such post-processing methods are relatively simple to train and implement in practice.

On the other hand, unrolled learned reconstruction operators aim to learn the reconstruction process itself. The goal is to learn a reconstruction operator that maps the measured data to high-quality image reconstructions. These methods incorporate the forward model into the network architecture and learn the reconstruction mapping directly from data. One prominent example is the Learned Primal-Dual (LPD) algorithm \cite{adler2018learned}, which unrolls a primal-dual optimization scheme into a trainable neural network architecture.

\section{Methods} \label{sec:methods}
We extend the self-supervised Noise2Inverse framework \cite{hendriksen2020noise2inverse} to the Learned Primal-Dual reconstruction method \cite{adler2018learned}. This yields a learned reconstruction operator that can be trained without access to ground-truth images, in contrast to supervised approaches such as U-Net–based denoisers \cite{gurrola2021residual}. In the following, we first describe the Noise2Inverse and Learned Primal-Dual methods separately, and then present their combination within a unified reconstruction framework.

\subsection{Noise2Inverse framework}
Noise2Inverse \cite{hendriksen2020noise2inverse} is a self-supervised framework, which utilizes independent subsets of projection angles in the training of the neural network, giving the possibility to train the networks without ground-truth images.

Considering the full sinogram data, it can be interpreted as a collection of sparsely sampled sub-sinograms. For example, if one has every even angle increment in one sub-sinogram, and every odd in the other, then one can combine these subsets into a more densely sampled one. Under the assumption that the measurement noise at different projection angles is mean-zero and statistically independent, the resulting sub-sinograms also contain independent noise realizations. 

Analytical reconstruction methods can then be applied separately to each sub-sinogram, producing reconstructions with independent noise components. Noise2Inverse utilizes these reconstructions as input--reference pairs for neural network training. Since the noise realizations vary independently between the input and target images, the network cannot consistently learn the noise and instead learns the underlying image structures. This splitting strategy can be implemented in different splits, which will be discussed later.

More precisely, let us consider a set of sinogram data $\{ \Tilde{g}_i \}_{i=1}^N$, where $\Tilde{g}_i$ denotes a noisy measurement 
\begin{equation*}
    \Tilde{g}_i = \mathcal{A}(f_i) + \varepsilon.
\end{equation*}
Here, the measurement operator $\mathcal{A}$ has dimensions $N_\alpha \times M$, where $N_\alpha$ denotes the number of projection angles and $M$ the number of detector elements (beams). Thus, each measurement $\Tilde{g}_i$ is obtained from $N_\alpha$ projection angles $\alpha_1, \alpha_2, \ldots, \alpha_{N_\alpha}$. The Noise2Inverse method proceeds as follows.

First, a hyperparameter $K$ is chosen, which defines an angle splitting\\  $\alpha_j, \alpha_{j+K}, \alpha_{j+2K}, \ldots$. The sinograms are then split into sub-sinograms $\Tilde{g}_{i,1}, \ldots, \Tilde{g}_{i,K}$, where each $\Tilde{g}_{i,j}$ contains measurements corresponding to every $K$th projection angle.
After splitting the data, sub-reconstructions are obtained as
\begin{equation*}
    \Tilde{f}_{i,j} = \mathcal{R}_j (\Tilde{g}_{i,j}), \quad j = 1, \ldots, K.
\end{equation*}
Here $\mathcal{R}_j$ is a reconstruction operator, and in this work, we use the filtered backprojection operator.

Next, the measurements are partitioned into a collection $\mathcal{J}$ of index sets $J \subset \{0,1, \ldots, N_\alpha\}$. For each $J \in \mathcal{J}$, we define the mean reconstruction
\begin{equation*}
    \Tilde{f}_{i,J} = \frac{1}{|J|} \sum_{j \in J} \Tilde{f}_{i,j} = \frac{1}{|J|} \sum_{j \in J} \mathcal{R}({\Tilde{g}}_{i,j}).
\end{equation*}
In this way, input–target pairs for training are constructed as $(\Tilde{f}_{i,J^C}, \Tilde{f}_{i,J})$, leading to the training problem
\begin{align}\label{eq:trained}
    \hat{\theta} &= \argmin_{\theta} \frac{1}{|\mathcal{J}|} \sum_{i=1}^N \sum_{J \in \mathcal{J}} \| \Lambda_\theta (\Tilde{f}_{i, J^C}) - \Tilde{f}_{i,J}\|^2_2 \\ &=  \argmin_{\theta} \frac{1}{|\mathcal{J}|} \sum_{i=1}^N \sum_{J \in \mathcal{J}} \| \Lambda_\theta(\RecOp( \Tilde{g}_{i, J^C})) - \Tilde{f}_{i,J}\|^2_2
    \\ &=  \argmin_{\theta} \frac{1}{|\mathcal{J}|} \sum_{i=1}^N \sum_{J \in \mathcal{J}} \| (\RecOp_\NNparam (\Tilde{g}_{i, J^C}) - \Tilde{f}_{i,J}\|^2_2
\end{align}
Note that here, in contrast to the original Noise2Inverse publication \cite{hendriksen2020noise2inverse}, we consider a learned reconstruction operator $\mathcal{R}_\NNparam$, since the aim is to utilize measurement data in the training.

To obtain a clean reconstruction $f^*_{i, \mathrm{out}}$, the network is evaluated as
\begin{equation}\label{eq:output}
    f^*_{i, \mathrm{out}} = \frac{1}{|\mathcal{J}|} \sum_{J \in \mathcal{J}} \Lambda_{\hat{\theta}} (\Tilde{f}_{i, J^C}),
\end{equation}
that is, predictions from different splits are averaged.

The key idea of Noise2Inverse is that the input and target images contain independent realizations of noise. As a result, the network learns to reconstruct the underlying signal while suppressing noise components. Intuitively, since the noise varies across the splits, it cannot be consistently learned by the network.

Let $\mathsf{g}$ and $\mathsf{f}$ denote random variables in the measurement space $Y$ and the image space $X$, respectively. The noisy measurements are given by
\begin{equation*}
    \Tilde{\mathsf{g}} = \mathcal{A} (\Tilde{\mathsf{f}}) + \varepsilon.
\end{equation*}
The corresponding sub-reconstructions are
\begin{equation*}
    \Tilde{\mathsf{f}}_{J^C} = \mathcal{R}_{J^C} (\Tilde{\mathsf{f}}_{J^C}), \text{ and } \Tilde{\mathsf{f}}_{J} = \mathcal{R}_{J} (\Tilde{\mathsf{f}}_{J}).
\end{equation*}
The trained network $\Lambda_{\hat{\theta}}$ from equation \ref{eq:trained} approximates the regression
\begin{equation}\label{eq:reg_approx}
    \NNOp_\NNparam^\dagger = \underset{\NNparam}{\mathrm{argmin}} \frac{1}{|\mathcal{J}|} \sum_{J \in \mathcal{J}} \mathbb{E}_{\mathsf{f}, \varepsilon} \big[ ||\NNOp_{\hat{\NNparam}} (\Tilde{\mathsf{f}}_{J^C}) - \Tilde{\mathsf{f}}_J ||^2_2 \big],
\end{equation}
which minimizes the expected prediction error.

To generalize this, let us randomize $J$, denoting it by $\mathsf{J}$. Now $\mathsf{J}$ takes uniformly random indices from the collection $\mathcal{J}$. This generalizes the equation \ref{eq:reg_approx} such that it becomes
\begin{equation}
    \NNOp_\NNparam^\dagger = \mathbb{E}_\mu \big[ ||\NNOp_{\hat{\NNparam}}(\Tilde{\mathsf{f}}_{\mathsf{J}^C}) - \Tilde{\mathsf{f}}_\mathsf{J} ||^2_2 \big],
\end{equation}
where $\mu$ is a joint measure of $\mathsf{f}, \varepsilon,$ and $\mathsf{J}$.

Before stating a proposition, let us define a clean sub-reconstruction of a clean measurement as
\begin{equation*}
    \mathsf{f}^*_\mathsf{J} = \mathcal{R}_\mathsf{J}(\mathsf{g}_\mathsf{J}).
\end{equation*}

\begin{proposition}\label{prop:exp_pred}
    Let $\Tilde{\mathsf{f}}_\mathsf{J}, \Tilde{\mathsf{f}}_{\mathsf{J}^C}, \mathsf{f}^*_\mathsf{J},$ and $\mu$ be defined as above. Let $\varepsilon$ be mean-zero and element-wise independent. Also, let $\mathcal{R}_J$ be a linear operator for all $J \in \mathcal{J}$. Then, for all learned reconstruction operators $\mathcal{R}_\NNparam: Y \to X$, we have
    \begin{equation}\label{eq:exp_pred}
        \mathbb{E}_\mu \big[ ||\mathcal{R}_\NNparam(\Tilde{\mathsf{f}}_{\mathsf{J}^C}) - \Tilde{\mathsf{f}}_\mathsf{J} ||^2_2 \big] = \mathbb{E}_\mu \big[ ||\mathcal{R}_\NNparam(\Tilde{\mathsf{f}}_{\mathsf{J}^C}) - \mathsf{f}^*_\mathsf{J} ||^2_2 \big] + \mathbb{E}_\mu \big[ ||\mathsf{f}^*_\mathsf{J} - \Tilde{\mathsf{f}}_\mathsf{J} ||^2_2 \big].
    \end{equation}
\end{proposition}
The proof for the same case with regression functions $h$ can be found in \cite{hendriksen2020noise2inverse}. The proposition shows that the expected prediction error decomposes into a supervised prediction error term and a noise variance term.

In practice, the splits are implemented by selecting subsets of projection angles. For experimental sinogram data, this can be done by indexing the sinogram array along the angular dimension. For simulated data, the splits can be implemented more accurately by defining corresponding forward operators for each subset.

\subsection{Effect of ReLU on Noise2Inverse training}
In this section, we 
analyze how the ReLU activation function influences training when used to enforce non-negativity in the Noise2Inverse setting. We use ReLU in all training, testing, and validation scenes, where in training we found that the loss is higher when ReLU is used, compared to the case where ReLU is absent. The proposition \ref{prop:exp_pred} does not include this case, but only considers the testing. Next, we derive the reasoning behind that.

Let us consider the $L_2$ loss functional
\begin{equation*}
     \mathcal{L}(\theta) = \frac{1}{|\mathcal{J}|} \sum_{J \in \mathcal{J}} ||\Lambda_\theta (\Tilde{f}_{J^C}) - \Tilde{f}_J||^2_2.
 \end{equation*}
Let the output of the network be a reconstructed image with 
an additive noise component $\varepsilon_s$. 
We denote this as $\Lambda_\theta(\tilde{f}_{J_C}) = \tilde{f}'_{J^C} + \varepsilon_s$, where $\varepsilon_s$ is a random variable representing residual noise. Then, using the definition of the ReLU function
\begin{equation*}
    \mathrm{ReLU}(x+y) = \max(x+y,0) = \frac{x+y+|x+y|}{2} \leq \frac{x+y+|x|+|y|}{2},
\end{equation*}
we obtain
\begin{align*}
    \tilde{f}_{J^C}' + \varepsilon_s &= \frac{\tilde{f}_{J^C}' + \varepsilon_s + \tilde{f}_{J^C}' + \varepsilon_s}{2}
    \\&\leq \frac{\tilde{f}_{J^C}' + \varepsilon_s + |\tilde{f}_{J^C}' + \varepsilon_s|}{2}
    \\&= \mathrm{ReLU}(\tilde{f}_{J^C}' + \varepsilon_s)
    \\&\leq \frac{\tilde{f}_{J^C}' + \varepsilon_s + |\tilde{f}_{J^C}'| + |\varepsilon_s|}{2}.
\end{align*}
With the inequality above, we can derive that
\begin{align*}
     \mathcal{L}(\theta) &= \frac{1}{|\mathcal{J}|} \sum_{J \in \mathcal{J}} ||\Lambda_\theta (\Tilde{f}_{J^C}) - \Tilde{f}_J||^2_2
     \\&\leq \frac{1}{|\mathcal{J}|} \sum_{J \in \mathcal{J}} ||\mathrm{ReLU}(\Lambda_\theta (\Tilde{f}_{J^C})) - \Tilde{f}_J||^2_2 = \mathcal{L}_{\mathrm{ReLU}}(\theta)
     \\&= \frac{1}{|\mathcal{J}|} \sum_{J \in \mathcal{J}} \bigg|\bigg| \frac{\Tilde{f}'_{J^C} + \varepsilon_s + |\Tilde{f}'_{J^C} + \varepsilon_s|}{2} - \Tilde{f}_J \bigg|\bigg|^2_2
     \\&\leq \frac{1}{|\mathcal{J}|} \sum_{J \in \mathcal{J}} \bigg|\bigg| \frac{\Tilde{f}'_{J^C}}{2} + \frac{\varepsilon_s}{2} + \frac{|\Tilde{f}'_{J^C}|}{2} + \frac{|\varepsilon_s|}{2} - \Tilde{f}_J \bigg| \bigg|^2_2
     \\&\leq \frac{1}{|\mathcal{J}|} \sum_{J \in \mathcal{J}} \bigg|\bigg| \frac{\Tilde{f}'_{J^C}}{2} + \frac{\varepsilon_s}{2} + \frac{|\Tilde{f}'_{J^C}|}{2} - \Tilde{f}_J \bigg| \bigg|^2_2 + \bigg| \bigg| \frac{|\varepsilon_s|}{2} \bigg| \bigg|^2_2.
\end{align*}
If we assume that the reconstruction is non-negative, then $|\Tilde{f}'_{J^C}| = \Tilde{f}'_{J^C}$, and the equation above reduces to
 \begin{equation}\label{eq:relu_loss}
     \mathcal{L}(\theta) \leq \frac{1}{|\mathcal{J}|} \sum_{J \in \mathcal{J}} \bigg|\bigg| \Tilde{f}'_{J^C} + \frac{\varepsilon_s}{2} - \Tilde{f}_J \bigg| \bigg|^2_2 + \bigg| \bigg| \frac{|\varepsilon_s|}{2} \bigg| \bigg|^2_2.
 \end{equation}
 From equation \ref{eq:relu_loss}, we observe that an additional constant term $\frac{|\varepsilon|}{2}$ is introduced into the loss. Furthermore, in the comparison between the output and the target reconstruction, the noise is effectively reduced by a factor of $\frac{1}{2}$.

\subsection{Learned Primal-Dual method}
Learned Primal-Dual (LPD) algorithm \cite{adler2018learned} is a data-driven approach for solving image reconstruction problems. It belongs to the class of learned iterative reconstruction algorithms, which are inspired by classical iterative optimization methods \cite{hauptmann2025learnediterativenetworksoperator}. The key idea is to use machine learning to learn the update rules at each iteration based on the current iterate.

The LPD method is based on adapting convolutional neural networks (CNNs) within the Primal-Dual Hybrid Gradient (PDHG) algorithm \cite{chambolle2011first}. In contrast to the classical PDHG algorithm, the proximal operators are replaced by learned proximals, that are neural networks.

\begin{algorithm}
    \caption{Learned Primal-Dual}
    \begin{algorithmic}[1]
        \State Initialize {$f_0 \in X, h_0 \in U$}
        \For {$i = 1, \ldots, I$}
            \State $h_{i} \gets \Gamma_{\theta^d_{i}} \big( h_{i-1}, \mathcal{A}f_{i-1}, g \big)$
            \State $f_{i} \gets \Lambda_{\theta^p_i} \big( f_{i-1}, \mathcal{A}^* h_{i} \big)$ 
        \EndFor
        \State \Return $f_I^{(1)}$
    \end{algorithmic}
    \label{algo:LPD}
\end{algorithm}

The LPD method is presented in Algorithm \ref{algo:LPD}. The algorithm consists of $2I$ learnable parameter spaces; $\Gamma_{\theta^d_{i}}$, corresponding to the dual updates, and $\Lambda_{\theta^p_i}$, corresponding to the primal updates. These components are implemented as neural networks and are separately optimized during training. In this work, the operator $\mathcal{A}$ denotes the ray transform and $\mathcal{A}^*$ its adjoint operator. The same operator $\mathcal{A}$ is also used as the forward model to generate the measurement data from the images in the first place. 

For the network architecture, we employ residual networks \cite{he2016deep}. The effect of different parameterizations is presented in Section \ref{sec:Results}.

\subsection{Learned Primal-Dual in the Noise2Inverse framework}
To combine the LPD reconstruction method with the N2I framework, several modifications are required, as the LPD architecture must be adapted to operate on split measurement data.

First, due to the data splitting in N2I, the forward operator $\mathcal{A}$ and its adjoint $\mathcal{A}^*$ must be adapted to the corresponding subsets of projection angles. This leads to a collection of split operators $\mathcal{A}_k$ and $\mathcal{A}_k^*$, where $k = 1, \ldots, K$ denotes the split index.

\begin{figure}
    \centering
    \includegraphics[width=0.6\linewidth]{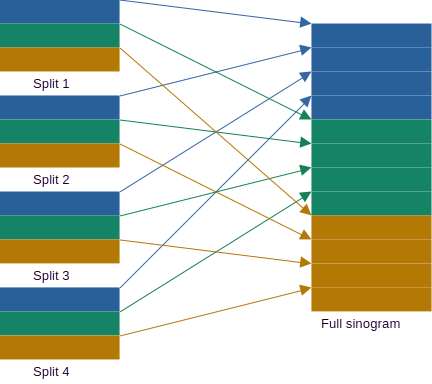}
    \caption{Visual explanation of the sinogram combining function (Algorithm \ref{alg:combine}). This example is for four splits.}
    \label{fig:explanation}
\end{figure}

Second, since the LPD algorithm operates on full sinograms in the dual space, the split sinograms must be recombined into a single sinogram. This is achieved using the sinogram combining function in Algorithm \ref{alg:combine}, which arranges the split sinograms according to their angular ordering, as illustrated in Figure \ref{fig:explanation}.

The overall workflow of the proposed method is described in Algorithm \ref{alg:N2I_LPD}. We initialize the primal and dual variables $f_0$, $g'$, and $h_0$, where $f_0$ is given by an FBP reconstruction, $g'$ is the full sinogram obtained from the splits, and $h_0$ is initialized as a zero tensor of the same size as $g'$.

Within each unrolled iteration, the split sinograms are first computed by applying the operators $\mathcal{A}_k$ to the current primal iterate, resulting in sub-sinograms stored in $g_f$. These are then combined into a full sinogram $g^*$ using the sinogram extension function. The dual variable $h_i$ is updated by passing $h_{i-1}$, $g^*$, and $g'$ through the dual network. This gives the full sinogram as an output.

After the dual network, an analytical reconstruction operator $\mathcal{A}_k^\dagger$, (for example, FBP) is applied to obtain $K$ sub-reconstructions $f_h$.
These are averaged to form a single reconstruction, which is then used to update the primal variable $f_i$ via the primal network, using $f_{i-1}$ and $f_h$. 

After $I$ unrolled iterations, the network produces an approximation of the reconstruction operator, 
and the final output is used for training and evaluation. This defines the learned reconstruction operator $\RecOp_\theta$ for the N2I-LPD method as outlined in Algorithm \ref{alg:N2I_LPD}.

The code for the implementation of this algorithm can be found from GitHub \cite{Sallinen_Noise2Inverse_Learned_Primal-Dual_2026}

\begin{algorithm}[ht]
        \caption{Noise2Inverse Learned Primal-Dual}
        \begin{algorithmic}[1]
            \State Initialize {$f_0 \in X^{j \times x_1 \times x_2}, g' \in Y^{y_1 \times y_2}, h_0 \in Y^{y_1 \times y_2}$}
            \For{$i = 1, \ldots, I$}
                \State $\{g_f^k\}_{k=1}^j \gets \mathcal{A}_k(f_{i-1})$ \textbf{for} $k=1,\ldots,j$
                \State $g^* \gets \mathrm{combine\_sinogram}(\{g_f^k\}_{k=1}^j)$
                \State $h_i \gets h_{i-1} + \Gamma_{\theta_i^d}(h_{i-1}, g^*, g')$
                \State $\{h_i^k\}_{k=1}^j \gets \mathrm{split}(h_i)$
                \State $\{f_h^k\}_{k=1}^j \gets \mathcal{A}_k^\dagger(h_i^k)$ \textbf{for} $k=1,\ldots,j$
                \State $f_h \gets  \frac{1}{j} \sum_{i=1}^{j} f^{i}_h$ 
                \State $f_i \gets f_{i-1} + \Lambda_{\theta_i^p}(f_{i-1}, f_h)$
            \EndFor
            \State $\RecOp_\mathbf{\theta}(g) \gets \text{ReLU}(f_I)$
        \end{algorithmic}
    \label{alg:N2I_LPD}
\end{algorithm}

\begin{algorithm}[ht]
    \caption{combine\_sinogram}
    \begin{algorithmic}[1]
        \State Initialize {$g_1 \in Y^{y_1 \times y_2}, g_2 \in Y^{k \times y_1 \times y_2}$}
        \For{$k = 0, \ldots, n$}
            \For{$j = 0, \ldots, y_1$}
                \State $g_1^{k+|k| \cdot j, y_2} \gets g_2^{k,j,y_2}$
            \EndFor
        \EndFor
        \State \Return $g_1$
    \end{algorithmic}\label{alg:combine}
\end{algorithm}

\section{Experiments} \label{sec:experiments}
In this section, we describe the CT data used in our experiments, how the corresponding measurements are simulated, and details of the neural network implementation.

\subsection{Walnut data set}\label{subsec:walnutdataset}
The data set consists of reconstructions of 42 different walnuts, acquired using a high-dose cone-beam CT setup \cite{der2019cone}. For each walnut, the data set provides a $501^3$ tensor of reconstructed cross-sectional images. These reconstructions are effectively noise-free and are therefore suitable for supervised learning, where they can serve as ground-truth images. For more information about the data set, we refer the reader to \cite{der2019cone}.

Since we consider a self-supervised setting in this work, we simulate the measurement data from these noise-free reconstructions. To do this, we use ODL \cite{adler2017operator}, a Python package. Specifically, we employ the \textit{FanBeamGeometry} and \textit{RayTransform} operators to generate sinograms from the reconstruction data. The fan-beam geometry used for each reconstruction is parameterized as follows: 512 projection angles uniformly distributed over $[0, 2\pi]$, and 496 detector elements (beams), with a source-to-object distance of 2.0 and an object-to-detector distance of 1.0. After generating the sinograms, we add $5\%$ zero-mean Gaussian noise to simulate a realistic measurement setting. Reconstructions are then obtained using the ODL function \textit{fbp\_op}.

Finally, we note that some slices in the data set contain mostly air. These slices are excluded from our experiments, as they are not relevant for the reconstruction task considered in this work.

\subsection{Network architecture and training}
\begin{table}[]
    \centering
    \begin{tabular}{|cccc|}
    \hline
    Model & \# Parameters & First channel & Depth \\
    \hline
    U-Net max & 7696706 & 64 & 4 \\
    U-Net min & 25522 & 16 & 2 \\
    \hline
    \end{tabular}
    \caption{Architectures of the U-Net models considered in this work. We include \textit{U-Net max} with 7,696,706 parameters and \textit{U-Net min} with 25,522 parameters. Here, \textit{First channel} denotes the number of channels in the first layer of the U-Net architecture, and \textit{Depth} refers to the total number of layers.}
    \label{tab:unet_parameters}
\end{table}

\begin{table}[]
    \centering
    \begin{tabular}{|ccccc|}
    \hline
    Model & \# Parameters & \# Channels & \# Layers & \# Unrolled iterations \\
    \hline
    LPD max & 578260 & 32 & 4 & 10 \\
    LPD min & 28730 & 16 & 2 & 5 \\
    \hline
    \end{tabular}
    \caption{Architectures of the Learned Primal-Dual models considered in this work. We include \textit{LPD max} with 578,260 parameters and \textit{LPD min} with 28,730 parameters.}
    \label{tab:lpd_parameters}
\end{table}

We use the U-Net \cite{ronneberger2015u} and ResNet \cite{he2016deep} architectures in our experiments, following the setups in Noise2Inverse \cite{hendriksen2020noise2inverse} and Learned Primal-Dual \cite{adler2018learned}. For the ResNet, we used the PReLU activation function between convolutional layers, following the original LPD implementation. In addition, we applied a final ReLU activation to both networks to enforce non-negativity of the output.

For the U-Nets, we used skip connections and $3 \times 3$ kernels for all 2D convolutions. For max-pooling and transposed convolutions, we used $2 \times 2$ kernels. Table \ref{tab:unet_parameters} shows how the number of channels in the first convolutional layer and the network depth affect the total number of parameters.

For the ResNets, we used $3 \times 3$ kernels for the 2D convolutions. The number of parameters was varied by changing the number of layers in the network, as well as the number of unrolled iterations in the algorithm. The corresponding configurations are summarized in Table \ref{tab:lpd_parameters}.

All neural networks were implemented using PyTorch \cite{paszke2019pytorch}. Training was performed for 50,000 iterations using the ADAM optimizer \cite{kingma2014adam} with a learning rate of 0.001, together with a cosine annealing learning rate schedule. We also applied gradient clipping with a maximum gradient norm of $1$, with respect to the $\ell^2$-norm. For the loss function, we used the standard $L^2$-loss in the training and evaluation, implemented using PyTorch's \textit{MSELoss}.

\subsection{Experimental setup}
In the Noise2Inverse framework, the data are partitioned into different bins (disjoint subsets). Following \cite{hendriksen2020noise2inverse}, we use a 3:1 split, where three subsets are used as input and one subset as comparison data. In our experiments, we also observed that a 1:1 split is not suitable for the N2I-LPD method, although the exact reason for this remains unclear. This splitting strategy effectively leads to sparse-angle reconstruction, since increasing the number of splits reduces the angular sampling in each sub-reconstruction.

For the network inputs, we use sinograms in methods that operate in the measurement domain, and corresponding FBP reconstructions for methods that operate in the image domain. At test time, the final reconstruction is obtained by evaluating the trained network on all four splits and averaging the resulting outputs.

\subsection{Comparison methods}
We consider several other reconstruction methods for comparison. The walnut data set contains ground truth images that were computed using FDK and accelerated gradient decent algorithm. For more information, see the original publication of the data set \cite{der2019cone}. FBP reconstructions are computed from the full noisy sinogram, and a classical Primal-Dual reconstruction \cite{chambolle2011first} with TV regularization.

For the FBP reconstruction, we use the ODL function \textit{fbp\_op}, which provides an approximate inverse of the \textit{RayTransform}. The parameters used are  \textit{padding=1}, \textit{filter\_type='Ram-Lak'}, and \textit{frequency\_scaling=1.0}.

The Primal-dual TV reconstruction algorithm is implemented as outlined in \cite{sidky2012convex}, using isotropic total variation. The forward operator is constructed using the ASTRA Toolbox \cite{van2016fast,van2015astra} along with the Spot linear operator toolbox \cite{spot2013}. Standard values for the primal-dual algorithm parameters were chosen: $\theta = 1$, $\sigma = \tau = 1/L$, where $L = ||(\mathcal{A},\nabla)^{\mathsf{T}}||_2$, and the forward operator $\mathcal{A}$ follows the fan-beam geometry outlined in \ref{subsec:walnutdataset}. In order to translate the ODL geometry parameters into a format suitable for ASTRA, a specific geometry dependent scaling factor\footnotemark[1] must be applied to the detector pixel size, the source-to-object distance, and the object-to-detector distance. A scaling factor value of $1.5735694$\footnotemark[2] was empirically determined to produce the best image registration between the ASTRA and ODL reconstructions. The finite differences in the total variation calculations were scaled by an empirically determined step size of $1/h = 85/(\textrm{scaling factor}) \approx 54$, which was found to best balance the magnitude of the data discrepancy and total variation terms in the primal-dual iterations.
\footnotetext[1]{url: \url{https://odlgroup.github.io/odl/_modules/odl/tomo/backends/astra_cuda.html}}
\footnotetext[2]{The theoretical scaling factor is given by $(\textrm{source-to-object distance} + \textrm{object-to-detector distance})/\textrm{source-to-object distance}$, which in our geometry equates to $1.5$. However, this value leaves a few pixels of registration error between the ASTRA and ODL reconstructions.}

\section{Results}\label{sec:Results}

\begin{figure}[t]
\centering
\includegraphics[width=\linewidth]{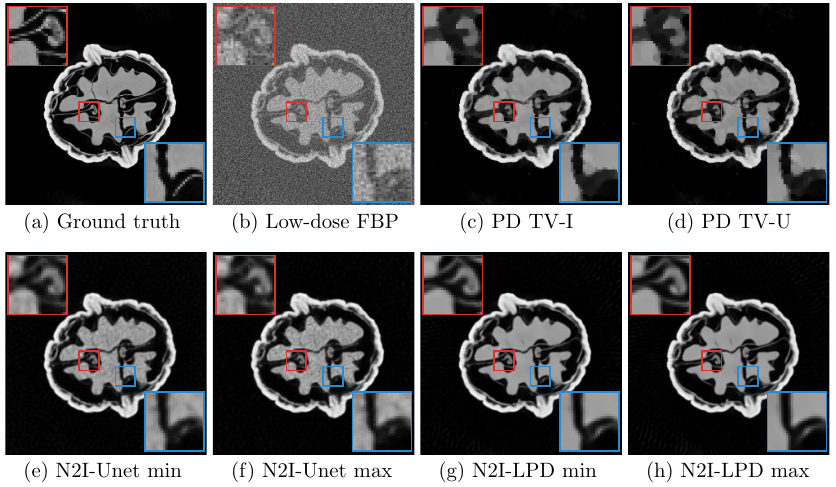}
\caption{Comparison of results for the Walnut data set.}
\label{fig:results_walnut}
\end{figure}

\begin{figure}[ht]
    \centering
     \begin{subfigure}[t]{0.49\textwidth}
         \centering
             \includegraphics[width=\textwidth]{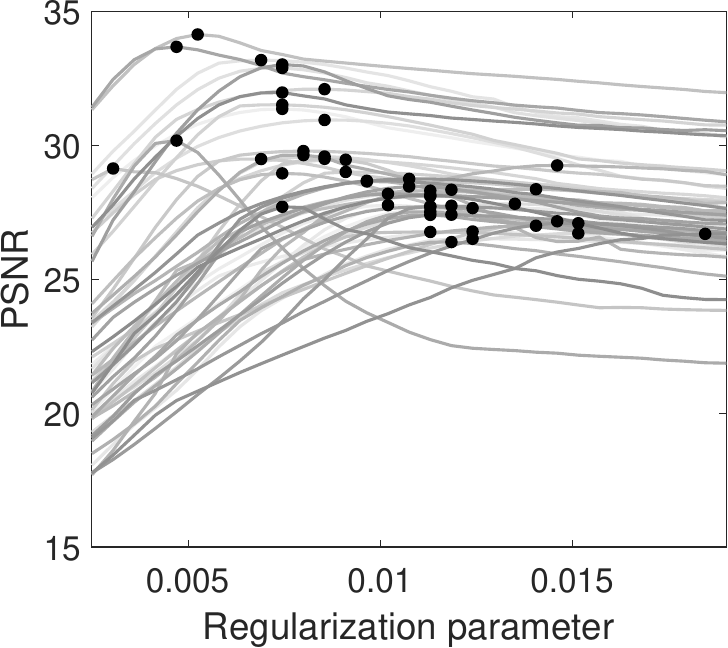}
         \caption{PD TV-I: individualized regularization parameter selection.}
         \label{fig:pdtv_regularization_all}
     \end{subfigure}
     \hfill
     \begin{subfigure}[t]{0.49\textwidth}
         \centering
         \includegraphics[width=\textwidth]{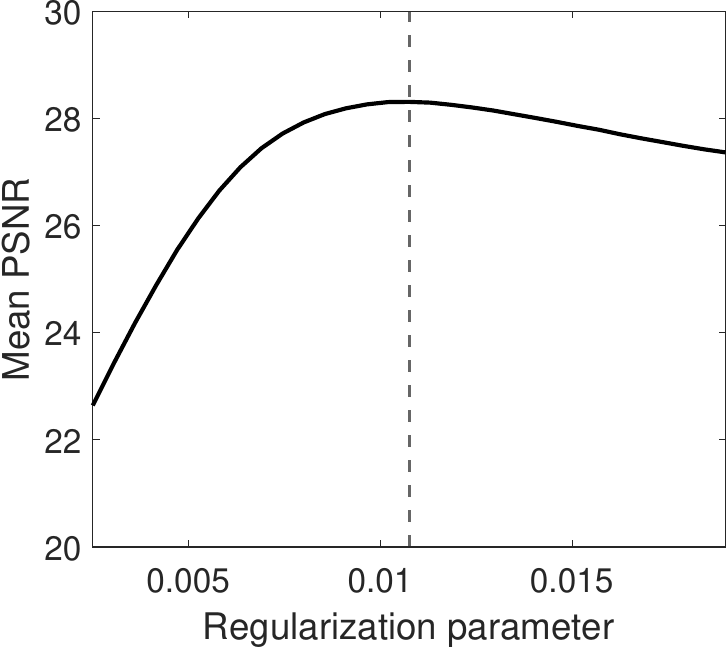}
         \caption{PD TV-U: uniform regularization parameter selection.}
         \label{fig:pdtv_regularization_mean}
     \end{subfigure}
    \caption{Regularization parameter choice for classical primal-dual with isotropic TV regularization. PSNR is computed for each choice of regularization parameter with respect to the ground truth image, for all 50 reconstructions. (\subref{fig:pdtv_regularization_all}) PSNR versus regularization parameter curves for all reconstructions. The maximum of each curve is indicated by the solid circles. (\subref{fig:pdtv_regularization_mean}) Mean PSNR value for a uniform choice of regularization parameter for all reconstructions. Dashed vertical line indicates the regularization parameter producing the maximal PSNR.}
    \label{fig:pdtv_regularization}
\end{figure}

In this section, we present the results of the proposed Noise2Inverse Learned Primal-Dual method. Both quantitative and qualitative evaluations are considered. For comparison, we include filtered backprojection, classical Primal-Dual methods with TV regularization, and N2I U-Net reconstructions. See Figure \ref{fig:results_walnut}. 

For the classical primal-dual method with TV regularization, reconstructions were generated for all of the 50 sinograms using 31 different regularization parameter values chosen uniformly from the interval $[0.0025,0.019]$. The PSNR with respect to the respective ground truth image was computed for each of these reconstructions. Two different methods for choosing the regularization parameter were considered: one where we choose the regularization parameters which maximize the PSNR for each individual reconstruction, and another where we choose a uniform regularization parameter which maximizes the mean PSNR of all the reconstructions, as seen in Figure~\ref{fig:pdtv_regularization}. The individualized regularization parameter choice in Figure~\ref{fig:pdtv_regularization_all} produces a best-case scenario for the average PSNR, whereas the uniform regularization parameter choice in Figure~\ref{fig:pdtv_regularization_mean} presents a more realistic value for the average PSNR of classical primal-dual total variation reconstructions, as ground truth images are not available in practice.

Table \ref{tab:results1} summarizes the quantitative results for all methods.
The results show that the N2I-LPD with the largest parameter configuration achieves the best performance, with an average PSNR of 29.62 dB. The N2I-LPD with the smallest parameter configuration and the TV-regularized PD method both achieve an average PSNR of approximately 29.00 dB. Both N2I U-Net models reach an average PSNR of 28.67 dB. In comparison, FBP performs significantly worse, with an average PSNR of 19.01 dB and SSIM of 0.20. The evaluation was performed on 50 slices that were not used during training.

For SSIM, the Primal-Dual TV method achieves the highest value (0.88), while the learned methods yield values between 0.68 and 0.76. All quantitative metrics were computed using the PSNR and SSIM implementations from PyTorch Ignite.  We verified the PSNR and SSIM values for the TV-regularized reconstructions using both PyTorch Ignite and MATLAB, obtaining consistent results.

\begin{table}[b]
    \centering
    \begin{tabular}{|ccc|}
    \hline
    Method & PSNR & SSIM \\
    \hline
    FBP & 19.01 & 0.20 \\
    PD TV-I & 29.00 & 0.88 \\
    PD TV-U & 28.31 & \textbf{0.89} \\
    N2I U-Net min & 28.67 & 0.68 \\
    N2I U-Net max & 28.67 & 0.70 \\
    N2I-LPD min & 29.39 & 0.75  \\
    N2I-LPD max & \textbf{29.62} & 0.76  \\
    \hline
    \end{tabular}
    \caption{Results for the walnut data set for all different methods used. Best values for PSNR and SSIM are highlighted in bold.}
    \label{tab:results1}
\end{table}

Next, we discuss the qualitative results presented in Figure \ref{fig:results_walnut}. Although the TV-regularized reconstructions achieve higher SSIM values, the visual differences in fine structural details compared to the N2I U-Net reconstructions are relatively small. The N2I U-Net results tend to preserve slightly sharper fine details, while the classical PD method produces stronger denoising effects. 

For the N2I-LPD reconstructions, both configurations exhibit improved preservation of fine details and smoother noise suppression compared to the other methods. In particular, they avoid the piecewise-constant appearance typically associated with TV regularization. Overall, the N2I-LPD methods achieve the best performance in terms of PSNR and provide visually improved reconstructions compared to the other methods.

\subsection{Generalizability}

\begin{figure}[t]
\centering
\includegraphics[width=\linewidth]{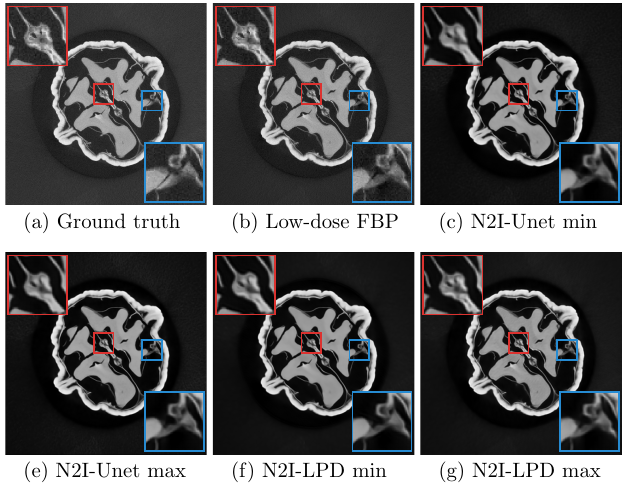}
\caption{Comparison of results for the generalized setting.}
\label{fig:results_gen}
\end{figure}

We also 
investigate the generalizability of the trained networks. To this end, we consider the cone-beam CT (CBCT) projections provided in the walnut data set, and construct a corresponding 2D-sinogram from them. This is done by using the projections acquired when the X-ray source is in the central position, as described in \cite{der2019cone}. For each projection, the center column is taken and assigned to the sinogram one column at a time. After constructing the full sinogram, the corresponding ODL geometry and operators are defined, after which the trained networks are evaluated on this previously unseen dataset.

Figure \ref{fig:results_gen} presents the reconstructions obtained by the aforementioned networks on previously unseen data. 
We note that the reference reconstructions provided in the Walnut dataset are computed using the FDK algorithm \cite{der2019cone}, and therefore do not exactly correspond to the FBP-based reconstruction setup considered in this work. Consequently, obtaining a perfectly matching ground-truth reconstruction for comparison is not possible in this setting. However, since the considered data are noise-free, the FBP reconstruction can still be used as a reliable qualitative reference. The ground truth reference reconstruction from the dataset is still included as the closest available approximation to the ground truth. While most structures are consistent across the reconstructions, Figure \ref{fig:results_gen} highlights regions where visible differences occur, indicated by the blue square.

All networks produce qualitatively similar reconstructions. However, in the highlighted region shown by the red square in Figure \ref{fig:results_gen}, N2I-LPD max provides the visually most accurate reconstruction. 

\begin{figure}[ht]
    \centering
    \includegraphics[width=0.5\linewidth]{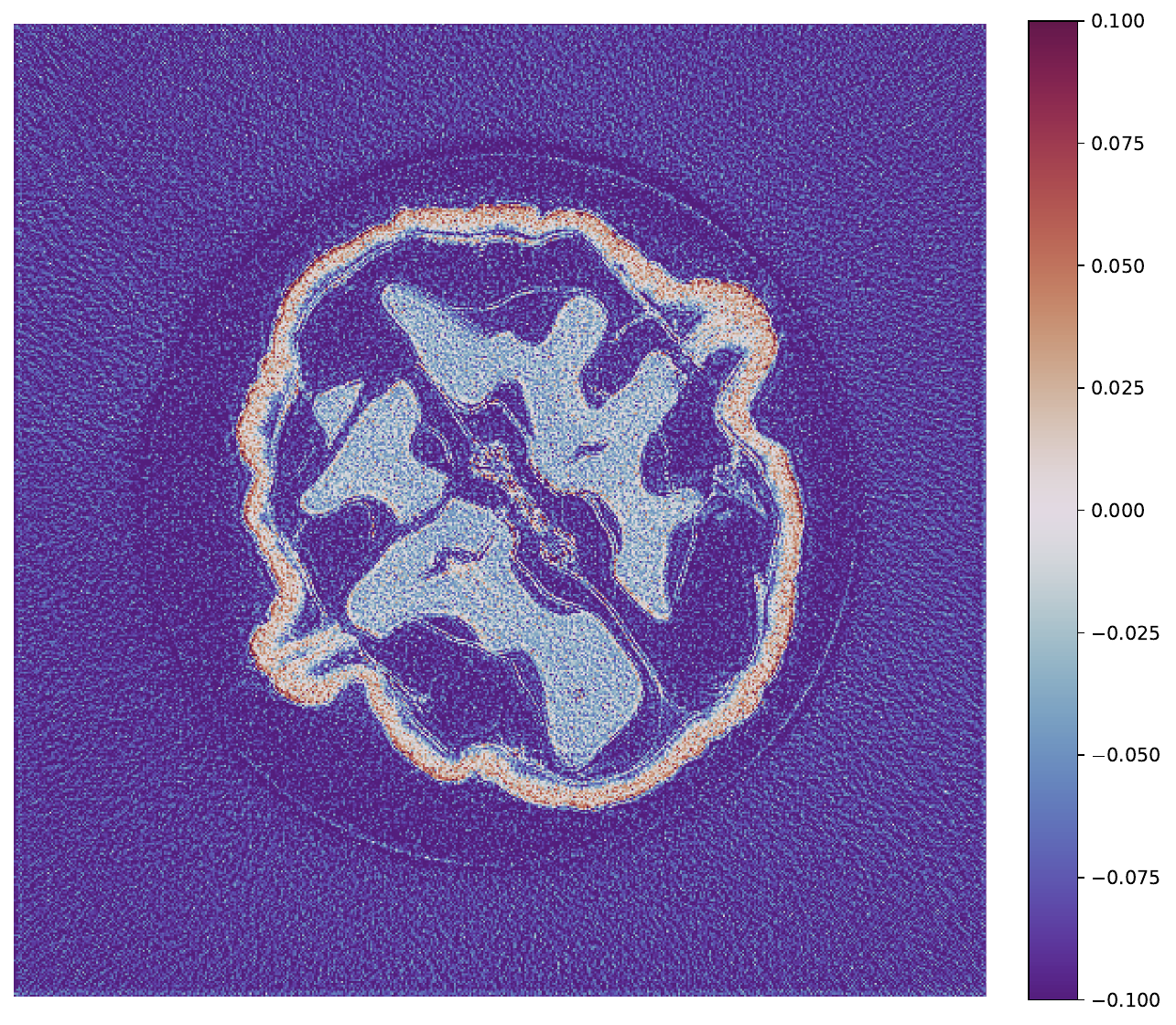}
    \caption{The difference map between the FBP reconstruction and the N2I-LPD max reconstruction in the generalized case.}
    \label{fig:diff_map}
\end{figure}

We also provide the difference map (see Figure \ref{fig:diff_map}) between the FBP reconstruction and the N2I-LPD max reconstruction, illustrating that no significant structural details are lost between the two reconstructions. Table \ref{tab:combination_results} shows that all neural networks evaluated in this generalized setting produce quantitatively similar results in terms of PSNR. These observations suggest that the networks generalize without introducing hallucinated structures into the reconstructions.

\begin{table}[ht]
    \centering
    \begin{tabular}{|l|ccc|}
    \cline{1-4}
    \multicolumn{1}{|c|}{Method} & N2I U-Net min & N2I U-Net max & N2I-LPD min  \\ \cline{1-4}
    N2I-LPD max & 36.73 & 36.81 & 38.26             \\ \cline{4-4}
    N2I-LPD min & 39.27 & 39.38 & \multicolumn{1}{|c}{}   \\ \cline{3-3}
    N2I U-Net max & 41.54 & \multicolumn{1}{|c}{}                    \\ \cline{1-2}
    \end{tabular}
    \caption{PSNR-values for all the combinations of the different network results in generalized setting used in this study.}
    \label{tab:combination_results}
\end{table}

\section{Discussion} 
\label{sec:discussion}
The results indicate that the proposed Noise2Inverse Learned Primal-Dual method achieves superior performance in terms of PSNR and visual reconstruction quality, while the TV-regularized Primal-Dual method attains the highest SSIM values. These findings suggest that incorporating learned reconstruction operators within self-supervised frameworks is a promising direction for low-dose and sparse-angle CT. These results further motivate several open questions regarding the proposed approach and its implementation.

First, in Algorithm \ref{alg:N2I_LPD}, the dual network operates on a full sinogram constructed from the split measurements. An alternative design would be to assign separate dual-space networks to each split. Such a modification could potentially allow the model to better exploit the statistical independence and zero-mean assumptions of the noise across the splits. A similar extension could be considered in the primal space, where instead of averaging back-projections prior to the network, separate networks could be applied to each reconstruction and their outputs combined.

Second, the choice of split ratio in the Noise2Inverse framework remains an open question. In this work, we follow \cite{hendriksen2020noise2inverse} and use a $1:4$ split, which was shown empirically to perform well. They divided the splitting strategy to be either $1:X$ or $X:1$. However, more general splitting strategies, such as $Y:X$ or $X:Y$ configurations, may affect both training stability and reconstruction quality. A systematic study of different split ratios could provide further insight into the behavior of the method.

Finally, an interesting question remains of how sparse of a setting could be used in the training to still achieve reasonable results. Increasing the number of splits effectively reduces the angular sampling in each subset, leading to more ill-posed sub-problems. Understanding how far this sparsification can be pushed while maintaining reconstruction quality is an important question for future work.

These observations highlight several directions for improving and extending the proposed N2I-LPD framework. In particular, they suggest that the design of the splitting strategy, the role of learned components in the primal and dual spaces, and the trade-off between sparsity and reconstruction quality are key factors that merit further investigation. Overall, the results demonstrate the potential of combining learned iterative reconstruction methods with self-supervised training strategies for challenging CT reconstruction settings.

\section{Conclusion} \label{sec:conclusion}
We introduce a novel reconstruction method, Noise2Inverse Learned Primal-Dual, which combines the N2I framework with the LPD algorithm, enabling training without ground-truth data. The proposed approach extends learned iterative reconstruction methods to a self-supervised setting. Our results demonstrate that the N2I-LPD method outperforms classical TV-regularized Primal-Dual reconstruction and N2I U-Net in terms of PSNR and visual reconstruction quality, while remaining competitive in SSIM. These findings highlight the potential of combining unrolled learned reconstruction operators with self-supervised training strategies for low-dose and sparse-angle CT imaging.

\section*{Acknowledgments}
This work has been supported in parts by the Research council of Finland (Project No. 359186, Flagship of Advanced Mathematics for Sensing Imaging and Modelling; Project No. 338408, Academy Research Fellow (AI-SOL); Project No. 370528, Academy project (AequiLoFi)) and by the Finnish Ministry of Education and Culture’s Pilot for Doctoral Programmes (Pilot project Mathematics of Sensing, Imaging and Modelling). S. Rautio also acknowledges support from the Väisälä Fund through the Finnish Academy of Science and Letters.

\newpage
\bibliographystyle{plain}
\bibliography{references.bib}

\end{document}